\documentclass[twocolumn,superscriptaddress,showpacs,prl,aps,floatfix,10pt]{revtex4-2}

\usepackage[dvips]{graphicx}
\usepackage{amsmath,amsfonts}
\usepackage{bm,dsfont}
\usepackage{dcolumn}

\def\ov#1{\overline{#1}{}}
\def\ul#1{\underline{#1}{}}

\def\ti#1{\tilde{#1}{}}\def\wti#1{\widetilde{#1}{}}

\def\bma{{\bm a}}\def\bmf{{\bm f}}\def\bmg{{\bm g}}\def\bmj{{\bm j}}
\def\bmk{{\bm k}}\def\bml{{\bm l}}\def\bmp{{\bm p}}\def\bmr{{\bm r}}
\def\bms{{\bm s}}

\def\bmA{{\bm A}}\def\bmB{{\bm B}}\def\bmC{{\bm C}}
\def\bmD{{\bm D}}\def\bmF{{\bm F}}\def\bmG{{\bm G}}
\def\bmK{{\bm K}}\def\bmL{{\bm L}}\def\bmP{{\bm P}}\def\bmR{{\bm R}}
\def\bmT{{\bm T}}\def\bmZ{{\bm Z}}

\def\cC{{\cal C}}

\def\cO{{\cal O}}\def\cS{{\cal S}}
\def\cU{{\cal U}}

\def\gb{\beta}\def\ggg{\gamma}\def\gd{\delta}\def\gvf{\varphi}
\def\gk{\kappa}

\def\gD{\Delta}\def\gG{\Gamma}\def\gL{\Lambda}\def\gO{\Omega}\def\gS{\Sigma}

\def\bgg{\bm{\gamma}}\def\bphi{\bm{\phi}}\def\bgz{\bm{\zeta}}

\def\ZZ{\mathds{Z}}

\def\re{\mathrm{e}}\def\ri{\mathrm{i}}
\def\Tr{\mathrm{Tr}\,}

\def\i{^{-1}}

\def\one{\mathds{1}}\def\RR{\mathds{R}}

\def\kost#1#2{(\,#1\,|\,#2\,)}
\def\ket#1{\!\mid~\!\!\!\!\!\!~{#1}~\!\!\rangle}

\def\mutm{{\mu t m}}\def\mut{{\mu t}}

\def\abs#1{\vert#1\vert}
\def\d={\buildrel \rm def \over =}

\DeclareMathOperator*{\Otimes}{\otimes}
\DeclareMathOperator*{\LProd}{\overleftarrow{\prod}}

\def\Briln{Brillouin}

\def\Grotendik{Grothendieck}

\def\Wannier{Wannier}

\def\Wyck{Wyckoff}
\begin{document}
\title{Closed Shell States in Periodic Compounds}

\author{Milan Damnjanovi\'c} 
\email[]{yqoq@rcub.bg.ac.rs}

\affiliation{NanoLab, Faculty of Physics, Uni Belgrade, Studentski trg 12, 11000 Belgrade, Serbia}
\altaffiliation{SANU,  Kneza Mihaila 36, 11000 Belgrade, Serbia }

\author{Ivanka Milo\v{s}evi\'{c}} 
\affiliation{NanoLab, Faculty of Physics, Uni Belgrade, Studentski trg 12, 11000 Belgrade, Serbia}

\begin{abstract}
Vanishing of the total angular momentum of the electrons occupying all orbitals of a closed shell in an atom is a textbook fact. Understanding the symmetry content of the atomic shell as irreducible representation of angular momentum, enables straightforward transfer of the notion to (translational or helically) periodic systems. More relevant generalizations naturally appear:  stratum shell is intermediate step to physically sound band representations, including elementary  and basic ones and connected components. We show that nontrivial determinant representation indicates stable topology of band in single colorless layer groups and obstructive limit in in single colorless line groups.

\end{abstract}

\maketitle

\section{Introduction}\label{SIntro}

Notion of the closed shell originates in atomic physics, where in the context of angular momentum (orbital, spin, total) it describes a set of single particle states of a multiplet $\ket{k,m}$ ($m=k,\dots,-k$) of the irreducible representation (IR) with weight $k$ of the angular momentum. Angular momentum of the electrons completely occupying such a multiplet (a closed shell) vanishes, and the total electronic angular momentum is effectively contributed only by the populated states from the opened (not fulfilled) multiplets. In fact, this means that the total electronic state, the Slater determinant $\ket{\Psi}$ of a closed shell is invariant under rotations. Clearly, if there are several fulfilled multiplets, they altogether have the same property, forming a closed shell in a bit broadened sense.

Surprisingly, there is no analogous notion for the crystalline systems; instead, as J. Birman~\cite{BIRMANJT1} formulated "crystal is normally considered to be in the $\gG^{1+}$ electronic state (single Slater determinant with all orbitals doubly occupied)", it is assumed that the total electronic state is invariant under the symmetry group. Here presented study, based on the common properties of the periodic (space, layer and line) groups reveals the complexity as well as the importance of this question.

Within single particle (tight-binding,  mean-field)  approximations (e.g.  Hartree-Fock, density functional) electron hamiltonian $H$ and a commuting with it representation $D(\bmG)$ of the symmetry group $\bmG$ are defined in the (single-particle) state space $\cS$.
The representation is reducible, and decomposes into the irreducible components $D^{(\mu)}(\bmG)$ with frequencies $f^\mu_D$:
\begin{equation}\label{EDdec}
  D(\bmG)=\sum_{\mu} f^\mu_DD^{(\mu)}(\bmG).
\end{equation}
Each eigenergy $E_{\mut}$ corresponds to the multiplet $\{\ket{\mutm}~|~m=1,\dots,\abs{\mu}\}$ such that $H\ket{\mutm}=E_{\mut}$. Each multiplet spans an irreducible subspace $\cS^{(\mu)}_{t}$, and for all $\mu$ and $t=1, \dots f^\mu$ these vectors form a stationary symmetry adapted basis of $\cS$.

A system of $N$ electrons occupies $N$ different single-particle states of $\cS$ due to Pauli principle.
This imposes a unique decomposition $\cS=\cC\oplus\cO\oplus\cU$ onto the maximal closed (fully occupied;  $\abs{\cC}\le N$) $\cC$, maximal unoccupied (empty) $\cU$ and the complemental open (partly occupied) $\cO$ subspaces invariant under $D(\bmG)$. This invariance implies that $D(\bmG)$ is reduced to the corresponding subrepresentations in these subspaces: $D(\bmG)=C(\bmG)+O(G)+U(\bmG)$. An $N$-electron state in a single-particle model is a Slater determinant of the filled states. Such $N$-particle Slater-determinants span the total state space, the antisymmetrized $N$-th tensor power $\cS^N_-$ of $\cS$. The group acts in $\cS^N_-$ as the antisymmetrized power $[D^N](\bmG)$ of $D(\bmG)$. The effective single particle state $\cS_{\mathrm{eff}}=\cC\oplus\cO$ yields the effective $N$-particle subspace $\cS^N_{\mathrm{eff}}=\ket{1,\dots,\abs{\cC}}\otimes\cO^{N-\abs{\cC}}_-<\cS^N_-$ spanned by the states matching the decomposition. It is straightforward to show that in this subspace the group is represented by
\begin{equation}\label{ENRep}
 D_{\mathrm{eff}}=\det{C}\ \,[O^{N-\abs{\cC}}].
\end{equation}
Note that the required invariance means that $\cC$, $\cO$ and $\cU$ contain only the whole multiplets, which prevents $\cC$ to be simply the span over the occupied states, with the complement $\cU$ (thus with no $\cO$).

It is well established that the ground (many-electron) state to a great extent determines properties of the condensed matter. Within the single particle prescription it emerges by filing the lowest energy $N$ eigenstates; the highest occupied one defines the Fermi level $E_\mathrm{F}$. Possible doping changes the available number of electrons (and $E_\mathrm{F}$), but once it is fixed $\cC$ and $\cO$ are well defined. Therefore, its  characterisation in terms of symmetry given by \eqref{ENRep} has the role of the total angular momentum in atomic or molecular physics.

Recall that a state $\ket{\Psi}$ of $N$ electrons is from the antisymmetrized tensor power $\cS^N_-$ of the single electron state space $\cS$ (infinite dimensional, but in molecular physics sometimes modeled by a finite dimensional subspace). If these electrons occupy states $\ket{\psi_i}$ ($i=1,\dots,N$) from $\cS$, then the total state is the Slater determinant expressed in terms of the permutations $\pi$ (with parity $\ti{\pi}$) from the symmetric group $S_N$, represented by the operators $\gD(\pi)$ gathered in the atisymmetrization operator  $A$:
\begin{subequations}\label{EState}\begin{eqnarray*}
\ket{\Psi}=\sqrt{N!}A\left(\Otimes\limits^N_{i=1}\ket{\psi_i}\right),&&
A=\tfrac1{N!}\sum_{\pi\in S_N}(-1)^{\tilde{\pi}}\gD(\pi),\ \ \ \ \ \\
\gD(\pi)\left(\Otimes\limits^N_{i=1}\ket{\psi_i}\right)&\d=&
\left(\Otimes\limits^N_{i=1}\ket{\psi_{\pi\i i}}\right).
\end{eqnarray*}\end{subequations}

\section{Determinant representations of crystal}\label{SBand}

Elements of a $\wp$-periodic (line, layer and space for $\wp=1,2,3$) group $\bmG$ are $g=\kost{G}{\bmg}$ (Koster's notation): $G$ is element of isogonal group $\bmP<O(3)$, while $\bmg$ is from the space $\RR^\wp$ containing lattice spanned by translational (or helical for $\wp=1$) invariant abelian subgroup $\bmT$, such that $\bmP=\bmG/\bmT$.

\Briln's zone $\mho$ is torus $T^\wp$ (elementary cell of the inverse lattice $\ov{\bmK}$), with points $\bmk$ (wave vectors) parameterizing IRs of $\bmT$. Representatives of the orbits (stars $\bmk^*$) of the group in \Briln's zone form the irreducible domain (ID) $\mho_\bmG$.
It is partitioned into strata $\mho^K_\bmG$ gathering equivalent representatives
(with conjugated, thus isomorphic, stabilizers $\bmF^\bmk\cong\bmF^K$). Each stratum is a manifold, with boundary (if not empty) consisting of strata with larger stabilizers. In particular, generic stratum occupies interior of ID, $\mho^{\mathrm{in}}_{\bmG}$, while the other, \emph{special} strata (planes, lines, points) are on its boundary. Irreducible representations of $\bmG$ (Appendix~\ref{AAIR})
associated to $\bmk$ are induced from the \emph{allowed} IRs of the stabilizer,
$D^{(K\bmk\gk)}(\bmG)=d^{(K\bmk\gk)}(\bmF^K)\uparrow\bmG$; accordingly,  $\mu=(K,\bmk,\gk)$,
where $\gk$ counts $\bmk$-independent (projective)  representations $d^{(K\gk)}(\bmF^K/\bmT)$ of the stratum little co-group; the dimension of $D^{(\mu)}$ is  $\abs{\mu}=\abs{\bmk^*}\abs{\gk}$.

Each determinant representation is one dimensional; besides irreducibility, this imposes two conditions to the associated $\bmk$ point: (a) its stabilizer is the whole group  (as $\abs{\bmk^*}=1$), and (b) it is simultaneously the allowed representation for which the factor system~\footnote{DO SADA ZAOBIDJENE PROJRKTIVNE; ako mora pomenuti ranije; ovde navesti referencu, zasto kod 1D nema netriv fac syst } (depending on $\bmk$ only) is (gauge) equivalent to the trivial (not projective) one. With $g=\kost{G}{\bmg}$ and  $g'=\kost{G'}{\bmg'}$ these conditions are:
\begin{subequations}\label{Egama}\begin{eqnarray}\label{EgamaFix}
&&G\bgg=\bgg+\bmK^{\bgg}_G,\quad \forall G\in\bmP\quad (\bmK^{\bgg}_G\in\ov{\bmK}),\\\label{EgamaRay}
&&\re^{\ri\bgg\cdot(G\bmg'-\bmg')}=\re^{\ri\bmK^{\bgg}_G\cdot\bmg'}=1,\quad \forall g,g'\in\bmG.
\end{eqnarray}\end{subequations}
Such, to be called  $\bgg$ points, form set $\mho^0_\bmG$. Hence, general form of determinant representation is: \begin{equation}\label{EDetRepGen}
 \ov{D}(g)\d=\det{D(g)}=\re^{\ri\bgg\cdot\bmg}d^{(\phi)}(G),\quad\bgg\in\mho^0_\bmG,
\end{equation}
where $d^{(\phi)}$ is one-dimensional representation of the isogonal group. Clearly, in all groups one $\gamma$ point is $\Gamma$-point ($\bmk=0$).  For abelian group $\bmG$ (pure translational, some layer and the first family line groups) all IRs are one-dimensional, and each $\bmk$ is $\bgg$-point: $\mho^0_\bmG=\mho_\bmG=\mho$. Otherwise, $\mho^0_\bmG$ is a proper lower dimensional subset of  $\mho_\bmG$. For symmorphic groups,  \eqref{EgamaRay} is automatically fulfilled for arbitrary fixed point (as then $\bmg$ is a lattice vector). Also, each fixed point within ID interior is a $\bgg$ point, as then $\bmK^{\bgg}_G=0$. Thus, only for the fixed points on the boundary of ID of nonsymmorphic groups \eqref{EgamaRay} is an effective restriction, which involves the whole group $\bmG$, and not only $\bmP$. The form \eqref{EDetRepGen} will be used for the forthcoming cases (details on derivation are in Appendices).

\section{Crystal shells}
Archetypic closed shell is that of electrons filling a multiplet of the angular momentum $\bmk$ (one of $\bml$, $\bms$ or $\bmj$) in $\cS$. For a fixed value $k$ of $\bmk$, states $\ket{\psi_i}$ are standard basis $\ket{k,m}$ ($m=-k,\dots,k$). For these $N=2k+1$ electrons, $K=0$ is the value of total angular momentum\cite{MES72} $\bmK=\sum_i\bmk_i$; common state, $\ket{\Psi}=\ket{K=0,M=0}$, is invariant under rotations. Indeed, arbitrary rotation $U$ acts on the states $\ket{k,m}$ by matrix $D^{(k)}(U)$ of the IR. But in $\cS^N_-$ it acts as $D(U)A=A\otimes^k_{m=-k}D^{(k)}(U)$, and $D(U)\ket{\Psi}=\det{D^{(k)}(U)}\ket{\Psi}=\ket{\Psi}$, since determinants of IRs of SU(2) are 1. Consequently, the total angular momentum is equal to the angular momentum of the electrons outside the closed shells.

Strict analogy with angular momentum establishes crystal shell as a multiplet of an IR of $\bmG$. The permutational-block-form of its matrices (Appendix~\ref{AdetIrPer}) yields sum $\bmk^o=\sum^{\abs{\bmk^*}}_{p=1}\bmk_p$ of the star vectors (phases of the blocks) in their determinants' form \eqref{EDetRepGen}):
\begin{equation}\label{EDetIR}
\ov{D}^{(K\bmk\gk)}\kost{G}{\bmg}=\re^{\ri\bgg\cdot\bmg}d^{(\phi_\gk)}(G),\quad \bgg=\abs{\gk}\bmk^o.
\end{equation}
This representation, depending on $\gk$  and $K$ (but not on $\bmk$) is not necessarily totally symmetric.

Some generalizations of crystal shell are natural for band theory; namely, then the state space carries a \emph{band representation}~\cite{CloizeauxBR,ZakBR,EvarestovBRVib,MICHELZAKebrs}
$D$, with $\bmk$-independent, constant along the associated strata, frequencies of the irreducible components  $f^{(K\bmk\gk)}_D=f^{(K\gk)}_D$:
\begin{equation}\label{EDdecC}
  D(\bmG)=\sum\nolimits_{K\gk}f^{(K\gk)}_D \sum\nolimits_{\bmk\in\mho^K_\bmG}D^{(K\bmk\gk)}(\bmG);
\end{equation}
actually, this property structures the spectrum of Hamiltonian in the form of bands consisted of \emph{band-patches} (parts over strata, having constant degeneracy) $E^t_{K\gk}(\bmk)$ (with $t=1,\dots,f^{(K\gk)}_D$). Each band-patch defines \emph{stratum shell} as the subspace assigned by \emph{stratum representation} $D^{[K\gk]}=\sum_{\bmk\in\mho^K_\bmG}D^{(K\bmk\gk)}$. Except for the point strata, or when $D^{(K\bmk\gk)}$ is unit representation, the corresponding determinant representation is mathematically ill defined continual product of determinants \eqref{EDetIR} over the stratum. Still,  coincident results of the
functional determinants method\footnote{Functional determinant method for an infinite dimensional operator $A$ uses $\det{A}=\re^{-\zeta'_A(0)}$, with the Riemann function $\zeta_A(x)=\Tr{A^{-x}}$. In the considered case $A=\sum_\bmk\re^{\ri(\abs{\gk}\bmk^o\cdot\bmg)}d^{(\gb)}(G)$ is diagonal infinite-dimensional direct sum (i.e. integral) of the determinants \eqref{EDetIR} over the domain $\mho^K_\bmG$. This gives \eqref{EDetStratum}, because $\zeta_A(x)=\int_{\mho^K_\bmG}\re^{-\ri(\abs{\gk}\bmk^o\cdot\bmg)x}d^{(\gb)^{-x}}(G)\mathrm{d}\bmk$.}
with the cellular decomposition starting with intuitive form
\begin{equation*}
\ov{D}^{[K\gk]}(g)=\re^{\ri\bgg^K_{\abs{\gk}}.\bmg}d^{(\Phi^K_\gk)}(G),\quad\bgg_K=\abs{\gk}\int_{\mho^K_\bmG}\bmk^o\mathrm{d}\bmk,
\end{equation*}
may justify both heuristics. Namely, $\phi_\gk$ from \eqref{EDetIR} is fixed along the stratum of integer measure $\abs{\mho^K_\bmG}$ ("number of points"), and $d^{(\Phi^K_\gk)}(G)=d^{(\phi_\gk)^{\abs{\mho^K_\bmG}}}(G)$ is continual power for a stratum of nonzero dimension.  Further, $d^{(\phi_\gk)}(G)$ is a root of 1 (represents an element of a finite group $\bmP$); hence, $d^{(\phi_\gk)}(G)=\re^{2\ri\pi z/Z}$ for some  (depending on $G$ and $\gk$, but not on $\bmk$) integers $Z$ (a divisor of the order of $G$) and $z\in[0,Z)$, giving $d^{(\Phi^K_\gk)}(G)=\re^{2\ri\pi z\abs{\mho^K_\bmG}/Z}$.
If $A$ is (topologically open) interior of  $\mho^K_\bmG$, it can be partitioned into $Z$ mutually equal open subdomains $A_i$ ($i=1,\dots,Z$) and boundaries between them. The contributions of the subdomains cancel and $d^{(\Phi^K_\gk)}(G)=\re^{2\ri\pi z\abs{B^K_1}/Z}$, where  union of all boundaries $B^K_1=\mho^K_\bmG\setminus(\bigcup^{Z}_iA_i)$ is of the lower (then stratum) dimension $b^K_1$. This can be calculated if $B^K_1$ is empty or finite set ($b^K_1=0$). Otherwise, the procedure is iterated (partition $B^K_1$  into $Z$ open subdomains gives again lower dimensional boundary $B^K_2$, etc.) until $n$ such that $b^K_n=0$ (examples in Fig.~\ref{FIDs2})
\begin{equation}\label{EDetStratum}
\ov{D}^{[K\gk]}(g)=\re^{\ri\bgg_K\cdot\bmg}\re^{ \tfrac{2\ri\pi z\abs{B^K_n}}{Z}},\quad\bgg_K=\abs{\gk}\int_{\mho^K_\bmG}\bmk^o_K\mathrm{d}\bmk;
\end{equation}
recall that $z$ and $Z$ are $G$- and $\gk$-dependent\footnote{Actually, only $|B^K_n|$ modulo $Z$ is relevant, and finer partitions do not change the result (e.g. simultaneous partition of all $A_i$ into the same number of open subsets enlarges $B^K_1$), meaning that common partition for all $G$ and $\gk$ can be found, as a characteristic of stratum. For example, $Z$ can be taken as the least common multiple of orders of all elements of $P$.}.

A band structure comprises of the glued  band patches; corresponding $D^{[K\gk]}$ are summed in the \emph{band representation} $D(\bmG)$. Its filled carrier space is final generalized notion, \emph{band shell}. Thus, \eqref{EDdecC} with \eqref{EDetStratum} gives:
\begin{equation}\label{ECdec}
\ov{D}(g)=\re^{\ri\bgg_D\cdot\bmg}\prod_{K\gk}d^{(\gb^K_\gk)^{f^{K\gk}_D}}(G),\ \bgg_D=\sum_{K\gk}f^{K\gk}_D\bgg_K.
\end{equation}
Note that this notion can be applied for connected structures of bands (otherwise open shell appearS);  each insulating band structure is a combination of band shells. Indeed, there is a set of  \emph{basic band representations} (BBRs) $B_i$ sufficient to  decompose (not always in a unique way) arbitrary band representation as $D=\sum_b c_b B_b$. Simultaneously,  the determinant representation of $D$ is factorized into determinant representations of BBRs (DBBRs): $\ov{D}=\otimes_b\ov{B}^{c_b}_b$, which shows how DBBRs determine  all band-shell representations.

\section{Discussion}

Symmetry enables to develop notion of the crystal closed shell hierarchically, from the spaces of IRs (like the atomic shell of angular momentum), over the strata shell to the connected components shells and basic band representations shells. Each of them is a determinant representation, an IR associated to one of the so called $\bgg$-points in BZ, with trivial factor system and fixed by the whole group (obviously including $\gG$-point); \eqref{EDetIR} shows that  the ray through the center of the star $\bmk^*$ contains $\gamma$-points of determinant representations of IRS  associated to $\bmk$. This result enable easy derivation of the higher band structure shells.

Determinant representations and closed shells appear in the context of Pauli antisymmetrization for fermions. This invokes half-integer representations (of double groups), although the determinant representation of even- or infinite-dimensional shell is an integer IR.  However, shells of single groups and determinant representations are useful in the analysis of influence of spin-orbit coupling, since the total representation is tensor product of the orbital shell with (two-dimensional) natural spin representation\cite{DLG}) of unit determinant. Consequently (see \eqref{EDetProd}), the orbital determinants  (e.g. of BBRs) are simply squared in total space. Besides these products, half-integer representations allow other band representations, which are related to spin-orbit interaction. We have calculated shells (of all types) determinants  of line and layer single and double groups.

Notice that the determinant representation is not a priori unit, as it is "normally considered"~\cite{BIRMANJT1}. Actually,  situation is quite interesting, and appears to be related to topology. To remind, there are two phenomena originating in topology.

Symmetry indicators\cite{WatanabeSI2} and topological quantum chemistry\cite{BernevigTQC} are based on the band representations (clearly, these generate all semiconducting shells).
To remind, \emph{elementary band representations}, $E=\{E_1,\dots,E_{\abs{E}}\}$,  are tabulated for space\cite{BernevigTQC}, layer\cite{YThessXX} and line\cite{LGEBR} groups. Within \Grotendik's extension to representation ring~\cite{WatanabeSI2,ShiozakiGomi}, they form abelian group $\bmA$ of \emph{atomic insulators} (with localized electrons), which is (normal) subgroup in the \emph{band structures group} $\bmB$, of the general band representations (satisfying only compatibility and degeneracy conditions). The factor group  of \emph{symmetry indicators} $\bm{SI}=\bmB/\bmA=\ZZ_1\otimes\cdots\otimes\ZZ_C$ is finite, with $C$ cyclic generators, and set of a \emph{stable} band representation $S=\{S_1,\dots,S_C\}$, each corresponding to a $\bm{SI}$ generator, complementing EBRs to the set of \emph{basic} band representations (BBRs). The elements of  $\bmB$ are combinations $D=\sum_e f^D_eE_e+\sum_s f^D_sS_s$, and band representations are those with non-negative IRs' frequencies $f^{K\gk}_D$. Within  $K$-theory, \emph{trivial} bands are non-negative combinations of EBRs ($f^D_e\ge0, f^D_s=0$), \emph{fragile topological bands} are expandable in EBRs ($f^D_s=0$) but with some negative EBRs' frequencies, while \emph{stable topological} bands necessarily contain stable components. For a given representation there may exist (topologically) different (e.g. in connectivity) band structures, which is related to Hamiltonian (dynamical details).

Another topological aspect is \emph{obstructed atomic limit} (OAL). In fact, in the usual models state space is inductive, with some orbitals associated to different atoms (orbits) sitting in {\Wyck}'s positions $x$. Corresponding orbitals span the spaces invariant under representation $d^x(\bmF_x)$ of the stabilizer $\bmF_x$. Summing the induced representations $d^x(\bmF_x)\uparrow\bmG$ over the orbit representatives, a BR $D$ is obtained, and it can be expanded over EBRs. It is expected that the electrons are localized ({\Wannier} centers) in the atoms. However, it may appear that this expansion is not unique, in which case the localization may be different (this is related to the change of Hamiltonian, i.e. to the change of phase). So, various expansions in EBRs of a BR is an indication of a possible OAL.

Namely, in single (colorless and grey) line groups, nonunit determinants appear exactly in the groups enabling OALs. In the layer groups trivial and fragile BRs are with unit determinants. In single groups nonunit shell corresponds to stable connected topological band; closed shell with several stable closed shells  may be unit. However, in double layer colorless groups, this is true only when  $\bm{SI}(\bmG)=\bm{SI}(\ti{\bmG})$ ($\ti{\bmG}$ denotes double group); in these cases (groups 2, 3, 6, 7, 49, 50, 51, 52, 65, 66, 73, 74 and 75) natural spin representation is reduced into two 1D components, and tensor product of stable BR of $\bmG$ or $\ti{\bmG}$ with any of these components gives a stable BR of $\wti{\bmG}$ and $\bmG$, respectively. When only $\bm{SI}(\ti{\bmG})$ is nontrivial, closed shells are unit, independent of the BR topology.

It should be noted that the non-unit closed shell representation may be important when correspond to  the ground states of DMI-systems. Then it is obviously indication of the vibronic instability.

\acknowledgments{This work is funded by Ministry of Science, Technological Development and Innovation, Projects 451-03-47/2023-01 and SANU-F93.}



%


\begin{thebibliography}{21}%
\makeatletter
\providecommand \@ifxundefined [1]{%
 \@ifx{#1\undefined}
}%
\providecommand \@ifnum [1]{%
 \ifnum #1\expandafter \@firstoftwo
 \else \expandafter \@secondoftwo
 \fi
}%
\providecommand \@ifx [1]{%
 \ifx #1\expandafter \@firstoftwo
 \else \expandafter \@secondoftwo
 \fi
}%
\providecommand \natexlab [1]{#1}%
\providecommand \enquote  [1]{``#1''}%
\providecommand \bibnamefont  [1]{#1}%
\providecommand \bibfnamefont [1]{#1}%
\providecommand \citenamefont [1]{#1}%
\providecommand \href@noop [0]{\@secondoftwo}%
\providecommand \href [0]{\begingroup \@sanitize@url \@href}%
\providecommand \@href[1]{\@@startlink{#1}\@@href}%
\providecommand \@@href[1]{\endgroup#1\@@endlink}%
\providecommand \@sanitize@url [0]{\catcode `\\12\catcode `\$12\catcode
  `\&12\catcode `\#12\catcode `\^12\catcode `\_12\catcode `\%12\relax}%
\providecommand \@@startlink[1]{}%
\providecommand \@@endlink[0]{}%
\providecommand \url  [0]{\begingroup\@sanitize@url \@url }%
\providecommand \@url [1]{\endgroup\@href {#1}{\urlprefix }}%
\providecommand \urlprefix  [0]{URL }%
\providecommand \Eprint [0]{\href }%
\providecommand \doibase [0]{https://doi.org/}%
\providecommand \selectlanguage [0]{\@gobble}%
\providecommand \bibinfo  [0]{\@secondoftwo}%
\providecommand \bibfield  [0]{\@secondoftwo}%
\providecommand \translation [1]{[#1]}%
\providecommand \BibitemOpen [0]{}%
\providecommand \bibitemStop [0]{}%
\providecommand \bibitemNoStop [0]{.\EOS\space}%
\providecommand \EOS [0]{\spacefactor3000\relax}%
\providecommand \BibitemShut  [1]{\csname bibitem#1\endcsname}%
\let\auto@bib@innerbib\@empty

\bibitem [{\citenamefont {Birman}(1962)}]{BIRMANJT1}%
  \BibitemOpen
  \bibfield  {author} {\bibinfo {author} {\bibfnamefont {J.~L.}\ \bibnamefont
  {Birman}},\ }\href {https://doi.org/10.1103/PhysRev.125.1959} {\bibfield
  {journal} {\bibinfo  {journal} {Phys. Rev.}\ }\textbf {\bibinfo {volume}
  {125}},\ \bibinfo {pages} {1959} (\bibinfo {year} {1962})}\BibitemShut
  {NoStop}%
\bibitem [{Note1()}]{Note1}%
  \BibitemOpen
  \bibinfo {note} {DO SADA ZAOBIDJENE PROJRKTIVNE; ako mora pomenuti ranije;
  ovde navesti referencu, zasto kod 1D nema netriv fac syst}\BibitemShut
  {NoStop}%
\bibitem [{\citenamefont {Messiah}(1972)}]{MES72}%
  \BibitemOpen
  \bibfield  {author} {\bibinfo {author} {\bibfnamefont {A.}~\bibnamefont
  {Messiah}},\ }\href@noop {} {\emph {\bibinfo {title} {Quantum Mechanics}}}\
  (\bibinfo  {publisher} {North-Holland, Amsterdam},\ \bibinfo {year}
  {1972})\BibitemShut {NoStop}%
\bibitem [{\citenamefont {des Cloizeaux}(1963)}]{CloizeauxBR}%
  \BibitemOpen
  \bibfield  {author} {\bibinfo {author} {\bibfnamefont {J.}~\bibnamefont {des
  Cloizeaux}},\ }\href@noop {} {\bibfield  {journal} {\bibinfo  {journal}
  {Phys. Rev. E}\ }\textbf {\bibinfo {volume} {129}},\ \bibinfo {pages} {554}
  (\bibinfo {year} {1963})}\BibitemShut {NoStop}%
\bibitem [{\citenamefont {Zak}(1980)}]{ZakBR}%
  \BibitemOpen
  \bibfield  {author} {\bibinfo {author} {\bibfnamefont {J.}~\bibnamefont
  {Zak}},\ }\href@noop {} {\bibfield  {journal} {\bibinfo  {journal} {Phys.
  Rev. Lett.}\ }\textbf {\bibinfo {volume} {45}},\ \bibinfo {pages} {1025}
  (\bibinfo {year} {1980})}\BibitemShut {NoStop}%
\bibitem [{\citenamefont {Evarestov}\ and\ \citenamefont
  {Smirnov}(1989)}]{EvarestovBRVib}%
  \BibitemOpen
  \bibfield  {author} {\bibinfo {author} {\bibfnamefont {R.~A.}\ \bibnamefont
  {Evarestov}}\ and\ \bibinfo {author} {\bibfnamefont {V.~P.}\ \bibnamefont
  {Smirnov}},\ }\href {https://doi.org/10.1002/pssb.2221520225} {\bibfield
  {journal} {\bibinfo  {journal} {physica status solidi (b)}\ }\textbf
  {\bibinfo {volume} {152}},\ \bibinfo {pages} {633} (\bibinfo {year}
  {1989})}\BibitemShut {NoStop}%
\bibitem [{\citenamefont {Michel}\ and\ \citenamefont
  {Zak}(2001)}]{MICHELZAKebrs}%
  \BibitemOpen
  \bibfield  {author} {\bibinfo {author} {\bibfnamefont {L.}~\bibnamefont
  {Michel}}\ and\ \bibinfo {author} {\bibfnamefont {J.}~\bibnamefont {Zak}},\
  }\href {https://doi.org/https://doi.org/10.1016/S0370-1573(00)00093-4}
  {\bibfield  {journal} {\bibinfo  {journal} {Physics Reports}\ }\textbf
  {\bibinfo {volume} {341}},\ \bibinfo {pages} {377 } (\bibinfo {year}
  {2001})},\ \bibinfo {note} {symmetry, invariants, topology}\BibitemShut
  {NoStop}%
\bibitem [{Note2()}]{Note2}%
  \BibitemOpen
  \bibinfo {note} {Functional determinant method for an infinite dimensional
  operator $A$ uses $\det {A}=\protect \mathrm {e}^{-\zeta '_A(0)}$, with the
  Riemann function $\zeta _A(x)=\protect \mathrm {Tr}\protect \,{A^{-x}}$. In
  the considered case $A=\DOTSB \sum@ \slimits@ _{\protect \bm {k}}\protect
  \mathrm {e}^{\protect \mathrm {i}(\vert \kappa \vert {\protect \bm
  {k}}^o\cdot {\protect \bm {g}})}d^{(\beta )}(G)$ is diagonal
  infinite-dimensional direct sum (i.e. integral) of the determinants \protect
  \eqref {EDetIR} over the domain $\mho ^K_{\protect \bm {G}}$. This gives
  \protect \eqref {EDetStratum}, because $\zeta _A(x)=\DOTSI \intop \ilimits@
  _{\mho ^K_{\protect \bm {G}}}\protect \mathrm {e}^{-\protect \mathrm
  {i}(\vert \kappa \vert {\protect \bm {k}}^o\cdot {\protect \bm
  {g}})x}d^{(\beta )^{-x}}(G)\protect \mathrm {d}{\protect \bm
  {k}}$.}\BibitemShut {Stop}%
\bibitem [{Note3()}]{Note3}%
  \BibitemOpen
  \bibinfo {note} {Actually, only $|B^K_n|$ modulo $Z$ is relevant, and finer
  partitions do not change the result (e.g. simultaneous partition of all $A_i$
  into the same number of open subsets enlarges $B^K_1$), meaning that common
  partition for all $G$ and $\kappa $ can be found, as a characteristic of
  stratum. For example, $Z$ can be taken as the least common multiple of orders
  of all elements of $P$.}\BibitemShut {Stop}%
\bibitem [{\citenamefont {Lazi\'{c}}\ \emph {et~al.}(2018)\citenamefont
  {Lazi\'{c}}, \citenamefont {Milivojevi\'{c}}, \citenamefont {Vukovi\'{c}},\
  and\ \citenamefont {Damnjanovi\'{c}}}]{DLG}%
  \BibitemOpen
  \bibfield  {author} {\bibinfo {author} {\bibfnamefont {N.}~\bibnamefont
  {Lazi\'{c}}}, \bibinfo {author} {\bibfnamefont {M.}~\bibnamefont
  {Milivojevi\'{c}}}, \bibinfo {author} {\bibfnamefont {T.}~\bibnamefont
  {Vukovi\'{c}}},\ and\ \bibinfo {author} {\bibfnamefont {M.}~\bibnamefont
  {Damnjanovi\'{c}}},\ }\href
  {http://stacks.iop.org/1751-8121/51/i=22/a=225203} {\bibfield  {journal}
  {\bibinfo  {journal} {Journal of Physics A: Mathematical and Theoretical}\
  }\textbf {\bibinfo {volume} {51}},\ \bibinfo {pages} {225203} (\bibinfo
  {year} {2018})}\BibitemShut {NoStop}%
\bibitem [{\citenamefont {Po}, \citenamefont {Vishwanath},\ and\ \citenamefont
  {Watanabe}(2017)}]{WatanabeSI2}%
  \BibitemOpen
  \bibfield  {author} {\bibinfo {author} {\bibfnamefont {H.~C.}\ \bibnamefont
  {Po}}, \bibinfo {author} {\bibfnamefont {A.}~\bibnamefont {Vishwanath}},\
  and\ \bibinfo {author} {\bibfnamefont {H.}~\bibnamefont {Watanabe}},\
  }\href@noop {} {\bibfield  {journal} {\bibinfo  {journal} {Nature
  Communication}\ }\textbf {\bibinfo {volume} {8}},\ \bibinfo {pages} {50}
  (\bibinfo {year} {2017})}\BibitemShut {NoStop}%
\bibitem [{\citenamefont {{B. Bradlyn and L. Elcoro and J. Cano and M. G.
  Vergniory and Z. Wang and C. Felser and M. Aroyo and A.
  Bernevig}}(2017)}]{BernevigTQC}%
  \BibitemOpen
  \bibfield  {author} {\bibinfo {author} {\bibnamefont {{B. Bradlyn and L.
  Elcoro and J. Cano and M. G. Vergniory and Z. Wang and C. Felser and M. Aroyo
  and A. Bernevig}}},\ }\href@noop {} {\bibfield  {journal} {\bibinfo
  {journal} {Nature}\ }\textbf {\bibinfo {volume} {547}},\ \bibinfo {pages}
  {298} (\bibinfo {year} {2017})}\BibitemShut {NoStop}%
\bibitem [{\citenamefont {Damnjanovi\'{c}}(2020)}]{YThessXX}%
  \BibitemOpen
  \bibfield  {author} {\bibinfo {author} {\bibfnamefont {M.}~\bibnamefont
  {Damnjanovi\'{c}}},\ }in\ \href@noop {} {\emph {\bibinfo {booktitle}
  {NANOTEXNOLOGY 2020, Thessaloniki, Greece, 4-11 July}}}\ (\bibinfo {year}
  {2020})\BibitemShut {NoStop}%
\bibitem [{\citenamefont {Milo\v{s}evi\'{c}}\ \emph {et~al.}(2020)\citenamefont
  {Milo\v{s}evi\'{c}}, \citenamefont {Dmitrovi\'{c}}, \citenamefont
  {Vukovi\'{c}}, \citenamefont {Dimi\'{c}},\ and\ \citenamefont
  {Damnjanovi\'{c}}}]{LGEBR}%
  \BibitemOpen
  \bibfield  {author} {\bibinfo {author} {\bibfnamefont {I.}~\bibnamefont
  {Milo\v{s}evi\'{c}}}, \bibinfo {author} {\bibfnamefont {S.}~\bibnamefont
  {Dmitrovi\'{c}}}, \bibinfo {author} {\bibfnamefont {T.}~\bibnamefont
  {Vukovi\'{c}}}, \bibinfo {author} {\bibfnamefont {A.}~\bibnamefont
  {Dimi\'{c}}},\ and\ \bibinfo {author} {\bibfnamefont {M.}~\bibnamefont
  {Damnjanovi\'{c}}},\ }\href {https://doi.org/10.1088/1751-8121/abba47}
  {\bibfield  {journal} {\bibinfo  {journal} {Journal of Physics A:
  Mathematical and Theoretical}\ }\textbf {\bibinfo {volume} {53}},\ \bibinfo
  {pages} {455204} (\bibinfo {year} {2020})}\BibitemShut {NoStop}%
\bibitem [{\citenamefont {Shiozaki}, \citenamefont {Sato},\ and\ \citenamefont
  {Gomi}(2017)}]{ShiozakiGomi}%
  \BibitemOpen
  \bibfield  {author} {\bibinfo {author} {\bibfnamefont {K.}~\bibnamefont
  {Shiozaki}}, \bibinfo {author} {\bibfnamefont {M.}~\bibnamefont {Sato}},\
  and\ \bibinfo {author} {\bibfnamefont {K.}~\bibnamefont {Gomi}},\ }\href
  {https://doi.org/10.1103/PhysRevB.95.235425} {\bibfield  {journal} {\bibinfo
  {journal} {Phys. Rev. B}\ }\textbf {\bibinfo {volume} {95}},\ \bibinfo
  {pages} {235425} (\bibinfo {year} {2017})}\BibitemShut {NoStop}%
\bibitem [{\citenamefont {Janssen}(1973)}]{TJANSSEN}%
  \BibitemOpen
  \bibfield  {author} {\bibinfo {author} {\bibfnamefont {T.}~\bibnamefont
  {Janssen}},\ }\href@noop {} {\emph {\bibinfo {title} {Crystallographic
  groups}}}\ (\bibinfo  {publisher} {North-Holland},\ \bibinfo {address}
  {Amsterdam},\ \bibinfo {year} {1973})\BibitemShut {NoStop}%
\bibitem [{\citenamefont {Evarestov}\ and\ \citenamefont
  {Smirnov}(2012)}]{EVARESTOVSiteSym}%
  \BibitemOpen
  \bibfield  {author} {\bibinfo {author} {\bibfnamefont {R.}~\bibnamefont
  {Evarestov}}\ and\ \bibinfo {author} {\bibfnamefont {V.~P.}\ \bibnamefont
  {Smirnov}},\ }\href@noop {} {\emph {\bibinfo {title} {Site Symmetry in
  Crystals: Theory and Applications}}},\ \bibinfo {series} {Springer Series in
  Solid-State Sciencesz}, Vol.\ \bibinfo {volume} {106}\ (\bibinfo  {publisher}
  {Springer},\ \bibinfo {year} {2012})\BibitemShut {NoStop}%
\bibitem [{\citenamefont {Nikoli\'{c}}\ \emph {et~al.}(2022)\citenamefont
  {Nikoli\'{c}}, \citenamefont {Milo\v{s}evi\'{c}}, \citenamefont
  {Vukovi\'{c}}, \citenamefont {Lazi\'{c}}, \citenamefont {Dmitrov\'{c}},
  \citenamefont {Popovi\'{c}},\ and\ \citenamefont
  {Damnjanovi\'{c}}}]{LaGSite}%
  \BibitemOpen
  \bibfield  {author} {\bibinfo {author} {\bibfnamefont {B.}~\bibnamefont
  {Nikoli\'{c}}}, \bibinfo {author} {\bibfnamefont {I.}~\bibnamefont
  {Milo\v{s}evi\'{c}}}, \bibinfo {author} {\bibfnamefont {T.}~\bibnamefont
  {Vukovi\'{c}}}, \bibinfo {author} {\bibfnamefont {N.}~\bibnamefont
  {Lazi\'{c}}}, \bibinfo {author} {\bibfnamefont {S.}~\bibnamefont
  {Dmitrov\'{c}}}, \bibinfo {author} {\bibfnamefont {Z.~P.}\ \bibnamefont
  {Popovi\'{c}}},\ and\ \bibinfo {author} {\bibfnamefont {M.}~\bibnamefont
  {Damnjanovi\'{c}}},\ }\href
  {https://doi.org/https://doi.org/10.1107/S205327332101322X} {\bibfield
  {journal} {\bibinfo  {journal} {Acta Cryst. A}\ }\textbf {\bibinfo {volume}
  {78}},\ \bibinfo {pages} {107} (\bibinfo {year} {2022})}\BibitemShut
  {NoStop}%
\bibitem [{\citenamefont {Nikoli\'{c}}\ \emph {et~al.}(2020)\citenamefont
  {Nikoli\'{c}}, \citenamefont {Milo\v{s}evi\'{c}}, \citenamefont
  {Vukovi\'{c}}, \citenamefont {Lazi\'{c}}, \citenamefont {Dmitrov\'{c}},
  \citenamefont {Popovi\'{c}},\ and\ \citenamefont {Damnjanovi\'{c}}}]{NLSite}%
  \BibitemOpen
  \bibfield  {author} {\bibinfo {author} {\bibfnamefont {B.}~\bibnamefont
  {Nikoli\'{c}}}, \bibinfo {author} {\bibfnamefont {I.}~\bibnamefont
  {Milo\v{s}evi\'{c}}}, \bibinfo {author} {\bibfnamefont {T.}~\bibnamefont
  {Vukovi\'{c}}}, \bibinfo {author} {\bibfnamefont {N.}~\bibnamefont
  {Lazi\'{c}}}, \bibinfo {author} {\bibfnamefont {S.}~\bibnamefont
  {Dmitrov\'{c}}}, \bibinfo {author} {\bibfnamefont {Z.~P.}\ \bibnamefont
  {Popovi\'{c}}},\ and\ \bibinfo {author} {\bibfnamefont {M.}~\bibnamefont
  {Damnjanovi\'{c}}},\ }\href {http://https://nanolab.group/} {\enquote
  {\bibinfo {title} {Nanolab site},}\ }\bibinfo {howpublished}
  {\url{https://nanolab.group}} (\bibinfo {year} {2020})\BibitemShut {NoStop}%
\bibitem [{\citenamefont {Damnjanovi\'{c}}\ and\ \citenamefont
  {Milo\v{s}evi\'{c}}(2010)}]{YILG}%
  \BibitemOpen
  \bibfield  {author} {\bibinfo {author} {\bibfnamefont {M.}~\bibnamefont
  {Damnjanovi\'{c}}}\ and\ \bibinfo {author} {\bibfnamefont {I.}~\bibnamefont
  {Milo\v{s}evi\'{c}}},\ }\href@noop {} {\emph {\bibinfo {title} {Line Groups
  in Physics: Theory and Applications to Nanotubes and Polymers}}},\ \bibinfo
  {series} {Lecture Notes in Physics}, Vol.\ \bibinfo {volume} {801}\ (\bibinfo
   {publisher} {Springer},\ \bibinfo {address} {Berlin},\ \bibinfo {year}
  {2010})\BibitemShut {NoStop}%
\bibitem [{\citenamefont {Damnjanovi\'{c}}\ \emph {et~al.}(1999)\citenamefont
  {Damnjanovi\'{c}}, \citenamefont {Milo\v{s}evi\'{c}}, \citenamefont
  {Vukovi\'{c}},\ and\ \citenamefont {Sredanovi\'{c}}}]{YCSYM}%
  \BibitemOpen
  \bibfield  {author} {\bibinfo {author} {\bibfnamefont {M.}~\bibnamefont
  {Damnjanovi\'{c}}}, \bibinfo {author} {\bibfnamefont {I.}~\bibnamefont
  {Milo\v{s}evi\'{c}}}, \bibinfo {author} {\bibfnamefont {T.}~\bibnamefont
  {Vukovi\'{c}}},\ and\ \bibinfo {author} {\bibfnamefont {R.}~\bibnamefont
  {Sredanovi\'{c}}},\ }\href@noop {} {\bibfield  {journal} {\bibinfo  {journal}
  {Phys. Rev. B}\ }\textbf {\bibinfo {volume} {60}},\ \bibinfo {pages} {2728}
  (\bibinfo {year} {1999})}\BibitemShut {NoStop}%
\end{thebibliography}

\bibliographystyle{aipnum4-2}

\appendix

\section{Induced representations}\label{AIR}

\subsection{Form of the induced representation}\label{AADetInd}
Let $\bmR=\{\bmr_p~|~p=1,2,\dots\}$ be an orbit of the symmetry group $\bmG$, i.e. all the sites are generated by the the action of the group transformations on a representative one, $\bmr_1$. There is a stabilizing subgroup $\bmF=\bmG_{\bmr_0}$ of $\bmG$, such that its elements $f$ fix $\bmr_0$, i.e. $f\bmr_1=\bmr_1$.
Accordingly, there is a coset decomposition of the group with a selection (\emph{transverzal}) $\bmZ=\{z_p~|~p=1,2,\dots\}$ of $Z=\abs{\bmG}/\abs{\bmF}=\abs{\bmR}$  coset representatives $z_p$, such that $z_p\bmr_1=\bmr_p$, i.e. each $z_p$ corresponds to the site $\bmr_p$. Each group element belongs to some coset of $\bmF$, meaning that it is uniquely factorized as $g=z(g)f(g)$. Then the site mapping $g\bmr_p=\bmr_{gp}$ corresponds to the equality
\begin{equation}\label{EGround}
gz_p=z_{\pi(g)p}f(g,p),
\end{equation}
and defines uniquely the integer $\pi(g)p$ (coset ordinal) and the stabilizer element $f(g,p)$. Associativity of group multiplication gives (for each $p$):
\begin{equation}\label{EGround2}
\pi(gh)=\pi(g)\pi(h),\ f(gh,p)=f(g,hp)f(h,p).
\end{equation}
The first property shows that $\pi(g)$ is a group homomorphism into the permutational group $S_Z$. It describes the permutational action of $\bmG$ over the sites. Matrices $E(g)=\sum_pE^{\pi(g)p}_p$
($E^p_q$ is matrix with elements $(E^p_q)_{ij}=\gd_{ip}\gd_{jq}$), are permutational,  with elements $E_{ij}(g)=\sum_p\gd_{i,\pi(g)p}\gd_{j,p}$ (column $j$ is zero, except that 1 is in the row $\pi(g)i$), and represent this action, assuming the site $\bmr_p$ is associated to the column $E^p$ (all zeros except 1 in the $p$-th row). These matrices, forming a permutational (ground) representation $E(\bmG)$ of the group,  relate group elements to permutations of the sites, identifying (as far as this action is considered) $g$ with a permutation $\pi(g)$ from permutational group $S_Z$. As a permutation each element $g$ obtains its parity $\ti{g}=\ti{\pi}(g)=\det E(g)$. In the following $\pi(g)p$ is shortened to $gp$ (e.g. $E^{gp}_p=E^{\pi(g)p}_p$).

Properties \eqref{EGround2} provide that for arbitrary representation $d(\bmF)$ of the stabilizer, the matrices
 \begin{equation}\label{EInd}
 D(g)=\sum^Z_{p=1}E^{gp}_p\otimes d(f(g,p))
\end{equation}
are representation of $\bmG$, called \emph{induced} and denoted as $D(\bmG)=d(\bmF)\uparrow\bmG$.

\subsection{Determinant of induced representation}\label{AADet}

For any representation $D$ of a group $\bmG$, the determinants $\det{D(\bmG)}$ preserve homomorphism, and form one-dimensional (thus irreducible) representation of $\bmG$. For direct sum and tensor product of representations
the determinant representations are
\begin{eqnarray}\label{EDetSum}
  \det({\oplus_i D_i}) &=&\prod\nolimits_i\det{D_i} \\\label{EDetProd}
  \det({\otimes_i D_i})&=&\prod\nolimits_i\det{^{\tfrac{\prod\nolimits_i\abs{D_i}}{\abs{D_i}}}}{D_i}.
\end{eqnarray}

Due to single nonzero block in each block-column and block-row, determinant of the matrix \eqref{EInd} factorizes, $\det{D(g)}=\det^{\abs{d}}{E(g)}\prod_{p}\det{d(f(g,p))}$, yielding:
\begin{equation}\label{EDetInd}
\det{D(g)}=\ti{g}^{\abs{d}}\prod^Z_{p=1}\det{d(f(g,p))}.
\end{equation}
Independence of the choices of the first element in a cycle and of the orbit representatives follows from the invariance of the determinant under the similarity transform.

\subsection{Determinant of the ground representation}\label{AADetGround}

As a permutation, element $g$ (i.e. $\pi(g)$) is partitioned into $C^g$ cycles, $g=g_1\cdots g_{C^g}$. Precisely, denote any $p$ as $s^1_1$, $gp$ as $s^1_2$, $g^2p=gs^1_2=s^1_3$, and so on, unless $g^{C^g_1}s^1_1=s^1_1$ for minimal integer $C^g_1$; then from complement of the subset $s^1=\{s^1_1,\dots,s^1_{C^g_1}\}$ another $p$ is chosen as $s^2_1$, and the procedure is repeated, unless the whole set $\{1,\dots,Z\}$ is exhausted in the last $C^g$-th cycle. In this way  the permutational action of $g$ is factorized into the independent permutations, cycles of $g$ of the length $C^g_c$, each of them permuting only the subset $s^c$; clearly, $\sum^{C^g}_{c=1}C^g_c=Z$ and $g^{C^g_1\cdots C^g_{C^g}}=\one$. If the sites are indexed such that the members of these subsets are one after another, then the matrix $E(g)$ is block-diagonal, with the cyclic permutational blocks of the dimensions $C^g_c$ on the diagonal, and
\begin{subequations}\label{ECyc}\begin{equation}\label{ECycZ}
\det{E(g)}=\prod_c\ti{g}_c=(-1)^{Z+C^g}.
\end{equation}
From $g^{C^g_c}s^c_1=s^c_1$ \eqref{EGround} implies $f(g^{C^g_c},s^c_1)=z\i_{s^c_1}gz_{s^c_1}$, and \eqref{EGround2} gives
$f(g^{C^g_c},s^c_1)=f(g^{C^g_c-1},s^c_2)f(g,s^c_1)$; iterations end up by:
\begin{equation}\label{ECycF}
z\i_{s^c_1}g^{C^g_c}z_{s^c_1}=f(g^{C^g_c},s^c_1)=f(g,s^c_{C^g_c})\cdots f(g,s^c_1).
\end{equation}\end{subequations}

\section{Periodic groups}
IRs of $\wp$-periodic group $\bmG$ are induced\cite{TJANSSEN} from the allowed IRs of the stabilizers of the $\bmk$-points in the irreducible domain $\mho_\bmG$. A brief reminder on their form is focused to analysis of their determinants with help of the above results.

\subsection{Irreducible representations}\label{AAIR}

The translational invariant abelian subgroup $\bmT$ of $\bmG$ (analogously, helical subgroup of line groups), as the direct product of $\wp$ (infinite) cyclic groups generated by translations for $\bma_i$ ($i=1,\dots,\wp$) has IRs
\begin{equation}\label{ETIR}
\gD^{(\bmk)}(\bma)=\re^{\ri\bmk\cdot\bma},\quad \bmk\in\mho=T^\wp.
\end{equation}
\Briln's zone, the set of different representations, is $\wp$ torus, since change of $\bmk$ for a vector of inverse lattice $\ov{\bmA}$ (with basis $\ov{\bma}_i$, such that $\ov{\bma}_i\cdot\bma_i=2\pi\gd_{ij}$), gives the same representation.  The action $g\bmk=G\bmk$ of $g=\kost{G}{\bmg}$ (effectively $G$ reduces to its restriction into the space spanned by lattice) from each $\bmk$ generates an orbit, the star $\bmk^*$. The orbits with conjugated stabilizers are gathered into strata; orbit representatives of the strata form representative strata $\mho^K_\bmG$ filling ID.

Whole $\bmT$ is invariant subgroup of the $\bmk$-point stabilizer $\bmF^\bmk$, and the factor group $\bmP^\bmk=\bmF^\bmk/\bmT$ is a subgroup of the isogonal group $\bmP=\bmG/\bmT$. Thus, the elements of $\bmF^\bmk$ are $f=\kost{F}{\bmf}=\kost{\one}{\bma_F}\kost{F}{\bphi}$, where $\kost{F}{\bphi}$ is arbitrary chosen $\bmT$-coset representative with element $F$ of $\bmP^\bmk$. Also,
any transversal $\{z_p=\kost{Z_p}{\bgz_p}\}$ of $\bmF^\bmk$ in $\bmG$ is determines the transverzal $Z_p$ of $\bmP^\bmk$ in $\bmP$: $\bmk^*$  vectors are $\bmk_p=Z_p\bmk$; conventionally, $\bmk=\bmk_1$ is from $\mho_\bmG$, and $\kost{Z_1}{\bgz_1}=\kost{\one}{0}$.

Irreducible representations~\cite{TJANSSEN} of $\bmG$ associated to $\bmk$ are the representations \eqref{EInd} induced from the stabilizer's $\bmF^\bmk$ \emph{allowed}  representations (irreducible, restricted to $\bmT$ decompose to a multiple of $\gD^{(\bmk)}(\bmT)$). The latter are
\begin{equation}\label{EDozvGk}
d^{(K\bmk\gk)}\kost{F}{\bmf}=\re^{\ri\bmk\cdot\bmf}d^{(K\gk)}(F);
\end{equation}
here $d^{(K\gk)}$ is projective IR of $\bmP^\bmk$ for the factor-system $\varphi_\bmk(F,F')=\re^{\ri\bmk\cdot(F\bphi'-\bphi')}$, which compensates fractional translations in the products $\kost{F}{\bphi}\kost{F'}{\bphi'}$):
\begin{equation}\label{ERay}
d^{(K\gk)}(F)d^{(K\gk)}(F')=
\re^{\ri\bmk\cdot(F\bphi'-\bphi')}d^{(K\gk)}(FF').
\end{equation}
As the strata of ID contain $\bmk$-points with the conjugated, thus isomorphic stabilizers, the isomorphic isogonal groups $\bmP^\bmk$ (projective) have the same IRs.
Finally, the allowed representation of stabilizer associates to a group element $g=\kost{G}{\bmg}=z_{gp}f_pz\i_p$ (here $f_p=\kost{F_p}{\bmf_p}$) the  matrix \eqref{EInd}:
\begin{equation}\label{ENPerIR}
D^{(K\bmk\gk)}(g)=\sum_{p=1}^{\abs{\bmk^*}}E^{gp}_{p}\otimes
\re^{\ri\bmk\cdot\bmf_p}d^{(K\gk)}(F_p).
\end{equation}
Notably, all the blocks vanish except $D^{(K\bmk\gk)}_{gp,p}(g)$. Since $F\i\bmk-\bmk$ is an inverse lattice vector, for arbitrary $\bma_{F'}$ holds $\re^{\ri\bmk\cdot(F\bphi'-\bphi')}=\re^{\ri\bmk\cdot(F\bmf'-\bmf')}$.

The IRs of double periodic group are of the same form, since $\bmT$ is their invariant subgroup, as well, and \Briln's zone, strata and ID are the same. Only the stabilizers are doubled, $\wti{\bmF}^\bmk$, with half-integer (extra) allowed representations; however, orbits and transverzals are the same, giving the same permutational structure of the block-matrices of the IRs~\cite{EVARESTOVSiteSym,DLG}.

\widetext
\subsection{Determinant of an irreducible representation of periodic groups: explicit derivation}\label{AdetIrPer}
For arbitrary group element $g=\kost{G}{\bmg}$ its powers are $g^n=\kost{G^n}{{\sum^n_{i=1}G^{n-i}\bmg}}=\kost{G^n}{{\sum^{n-1}_{i=0}G^i\bmg}}$.
The product of several elements, denoted as $\LProd^n_{i=1}g_i\d=g_n\cdots g_1$, is
$\LProd^n_{i=1}\kost{G_i}{\bmg_i}=\kost{\LProd^n_{i=1}G_i}{\sum^n_{i=1}\LProd^n_{t=i+1}G_t\bmg_i}$.
For $z_p=\kost{Z_p}{\bgz_p}$ and $z_{cj}=\kost{Z_{cj}}{\bgz_{cj}}$ it follows
$z\i_pg^iz_p=\kost{Z\i_pG^iZ_p}{Z\i_p(G^i\bgz_p-\bgz_p+\sum^{i-1}_{j=0}G^j\bmg)}$ and
$z\i_{c1}g^{C^c_g}z_{c1}=\kost{F^c}{\bmf^c}=\kost{Z\i_{c1}G^{C^c_g}Z_{c1}}{Z\i_{c1}(G^{C^c_g}\bgz_{c1}-\bgz_{c1}+\sum^{{C^c_g}-1}_{j=0}G^j\bmg)}$.
Substututing \eqref{ENPerIR} in \eqref{EDetInd}, with $\bmk^o\d=\sum^{\abs{\bmk^*}}_{p=1}\bmk_p$ being the sum of the star wave vectors (and sum over cycle $\bmk^o_c\d=\sum^{C^g_c}_{j=1}\bmk^c_j$), one gets:

\begin{eqnarray*}
\det{D^{(K\bmk\gk)}(g)}&=&(-1)^{\abs{\gk}(\abs{\bmk^*}+C^g)}
  \re^{\ri\abs{\gk}\bmk\cdot\sum^{\abs{\bmk^*}}_{p=1}\bmf_p}
  \prod^{\abs{\bmk^*}}_{p=1}\det{d^{(K\gk)}(F_p)}\\
&=&(-1)^{\abs{\gk}(\abs{\bmk^*}+C^g)}\prod^{C^g}_{c=1}
  \det{d^{(\bmk,\gk)}\kost{Z\i_{c1}G^{C^g_c}Z_{c1}}{Z\i_{c1}(G^{C^g_c}\bgz_{c1}-\bgz_{c1}+\sum^{{C^g_c}-1}_{j=0}G^j\bmg)}}\\
&=&(-1)^{\abs{\gk}(\abs{\bmk^*}+C^g)}\prod^{C^g}_{c=1}
\det{\re^{\ri\bmk\cdot Z\i_{c1}(G^{C^g_c}\bgz_{c1}-\bgz_{c1}+\sum^{C^g_c-1}_{j=0}G^j\bmg)}d^{(K\gk)}(Z\i_{c1}G^{C^g_c}Z_{c1})}\\
&=&(-1)^{\abs{\gk}(\abs{\bmk^*}+C^g)}
\re^{\ri\abs{\gk}\sum^{C^g}_{c=1} Z_{c1}\bmk\cdot(G^{C^g_c}\bgz_{c1}-\bgz_{c1}+\sum^{C^g_c-1}_{j=0}G^j\bmg)}
\prod^{C^g}_{c=1}\det{d^{(K\gk)}(Z\i_{c1}G^{C^g_c}Z_{c1})}\\
\\
&=&(-1)^{\abs{\gk}(\abs{\bmk^*}+C^g)}
\re^{\ri\abs{\gk}\sum^{C^g}_{c=1} (\bmK^g_{c1}\cdot\bgz_{c1}+\bmk_{c1}\cdot\sum^{C^g_c-1}_{j=0}G^j\bmg)}
\prod^{C^g}_{c=1}\det{d^{(K\gk)}(Z\i_{c1}G^{C^g_c}Z_{c1})}.
\end{eqnarray*}

Lattice translations $g=\kost{\one}{\bma}$ fix each $\bmk$, being thus identity permutation $\pi(g)=\one$. Thus, they have $C^g=\abs{\bmk^*}$ cycles ($p=c$) of the lengths $c^g_p=1$; further, $gz_p=\kost{Z_p}{\bgz_p}\kost{\one}{Z\i_p\bma}$ $z_{gp}=z_p$, $f_p=z\i_{gp}gz_p=\kost{\one}{Z\i_p\bma}$, i.e. $F_p=\one$ and $\bmf_p=Z\i_p\bma$, giving:
\begin{equation}\label{EDetk}
\det{D^{(K\bmk\gk)}\kost{\one}{\bma}}=(-1)^{2\abs{\gk}\abs{\bmk^*}}
  \re^{\ri\abs{\gk}\sum^{\abs{\bmk^*}}_{p=1}Z_p\bmk\cdot\bma}
  \prod^{\abs{\bmk^*}}_{p=1}\det{\one}
  =\re^{\ri\abs{\gk}\bmk^o\cdot\bma}
\end{equation}
\endwidetext

\subsection{$\bgg$-points}
This reveals that the determinants form a representation associated to the $\bgg$-point $\abs{\gk}\bmk^o$, which allows
one-dimensional representations. For the pure translational and other Abelian groups all IRs are one-dimensional, and each $\bmk$ is a $\ggg$-point:  $\mho^0_\bmG=\mho_\bmG=\mho$. In other groups $\mho^0_\bmG$ is discrete subset of  $\mho_\bmG$.

According to \eqref{ENPerIR}, the one-dimensionality implies
that each $\bgg$ is a fixed point for the whole group (otherwise the induction gives larger dimension),
$\bmF^{\bgg}=\bmG$ (and $\bmP^{\bgg}=\bmP$), as well as that the allowed representation of $\bmP$ is an ordinary representation. Therefore $\mho^0_\bmG$ belongs to the maximally symmetric (regarding stabilizers) strata of ID, which are of the minimal dimension, and at the boundary of all other strata.
Since $\bgg$ is fixed (up to the inverse lattice) by the whole isogonal group, it obeys \eqref{EgamaFix}.
Obviously, $\gG$-point, $\bgg=0$, is a solution. For $\bgg$ in the interior of $\mho_\bmG$, only $\bmK^{\bgg}_P=0$ is possible; thus, the fixed points of $\bmP$ in  $\mho_\bmG$, besides those fixed by the Euclidean action (in the lattice subspace), may be only on the boundary of $\mho$.

Still, there are fixed points with no one-dimensional representations. In fact, as irreducible, such a representation is of the form \eqref{EDozvGk}, with $\bmP^{\bgg}=\bmP$, and the factor system satisfying \eqref{ERay}. On the other side, one-dimensional representations are projectively equivalent to the unit representation. Taking in \eqref{EDozvGk}  $d^{(K\gk)}(P)=1$ (as a projective unit representation), and multiplying each of them by the corresponding matrix $u(P)=\gD^{(\gb)}(P)$ of one-dimensional (ordinary) representation of $\bmP$, the new multipliers $\gvf'_{\bgg}(P,P')=
u(P)u(P')\gvf_{\bgg}(P,P')/u(PP')$ must be trivial. Thus, fixing a representative $p=\kost{P}{\bmp}$ for each element $P$ of $\bmP$, and applying \eqref{ERay}, in the view of the assumption that $\gD^{(\gb)}(\bmP)$ is a representation, the sufficient condition for a fixed point to be a $\bgg$-point becomes \eqref{EgamaRay}.
Note the independence of $\gD^{(\gb)}(\bmP)$, i.e. all one-dimensional representations are simultaneously allowed.
For symmorphic groups this is automatically fulfilled (as then $\bmp$ is a lattice vector), as well as within interior of $\mho$ (then $\bmK^{\bgg}_P=0$).

To resume, $\bgg$-points are points fixed by the whole group, with one-dimensional associated IRs; they satisfy \eqref{Egama}. The fixed  points within interior of ID are automatically $\bgg$-points (actually, in this case the corresponding $d^{(K\gk)}(\bmP^\bmk)$ in \eqref{EDozvGk} are ordinary representations). As for symmorphic groups, all fixed points are $\ggg$-points. Thus, only for the fixed points on the boundary of ID for nonsymmorphic groups \eqref{EgamaRay} is an effective restriction, depending not only on $\bmP$, but on the whole group $\bmG$ (only they can differ for the groups with the same $\bmP$). Finally, when \eqref{Egama} are fulfilled, the determinant representation is
\begin{equation}\label{EDetIR1}
\det{D^{(K\bmk\gk)}(g)}=\re^{\ri\bgg\cdot\bmg}d^{(\gb)}(G),\quad \bgg=\abs{\gk}\bmk^o,
\end{equation}
where $d^{(\gb)}(\bmP)$ is one of the one-dimensional (ordinary) representation of the isogonal group.

\subsubsection{Layer groups}

\begin{figure}[hbt]\begin{center}
\includegraphics[width=0.5\textwidth]{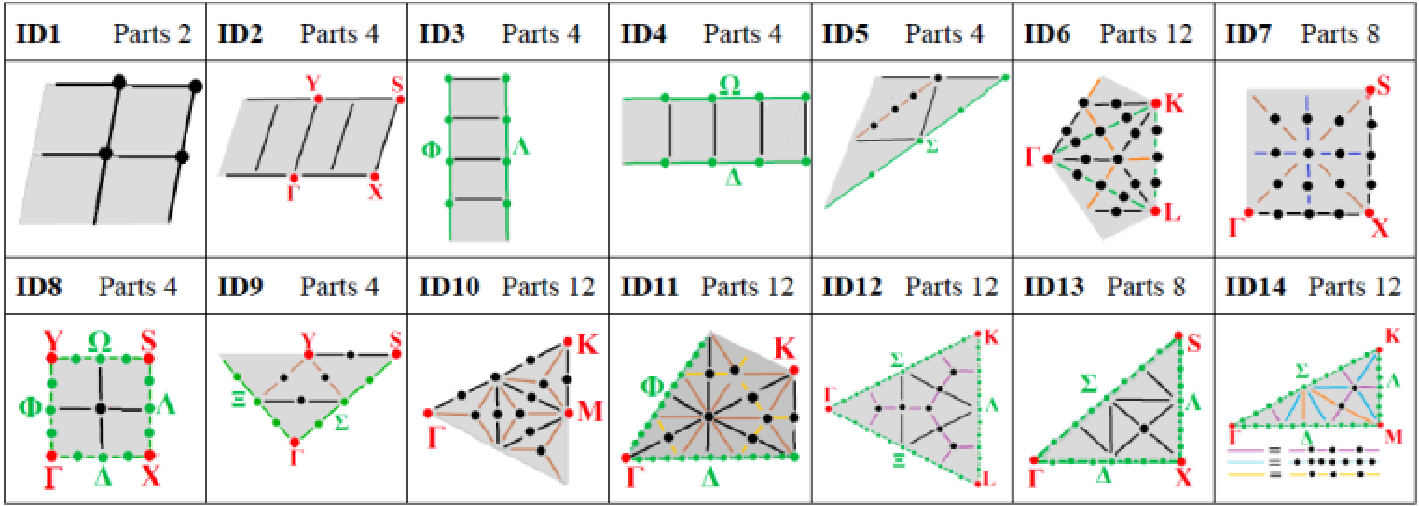}
\caption{\label{FIDs2}For each irreducible domain (as given in Ref.~\onlinecite{LaGSite,NLSite}, the cellular division into $d$  equivalent parts is presented. Point strata are red, and line ones are green (with green included edges and remaining points). Interior is generic stratum (grey, with included edges emphazized by black lines) is divided in with remaining lines (the equivalent ones are in the same color), which are then divided into open intervals and remaining points.}
\end{center}\end{figure}

\squeezetable
\begin{table}
\caption{\label{TIDs}For each irreducible domain ID (not bolded if for colorless groups only), (standard~\cite{NLSite}) of layer groups the isogonal group $\bmP$, $n$ orders of the elements $n$, division order (LCM of the orders of the double group generators) $d$, and number $c$ of contributing points for strata $S$ (underlined if $\bgg$-point) are listed.}
\begin{tabular}{|rcc|cl|}
  \hline
  ID&$\bmP$&$n$&$d$&$cS$ and $c\ul{\bgg}$ \\     \hline
   1           &$\bmC_1$  &1  & 4&$0\ul{G}$\\
   $\mathbf{2}$&$\bmC_2$  &2  & 4&$1\ul{\gG},1\ul{X},1\ul{Y},1\ul{S},0G$\\
   3           &$\bmD^2_1$&2  & 4&$0\ul{\Phi},0\ul{\gL},0G$\\
   4           &$\bmD^1_1$&2  & 4&$0\ul{\gD},0\ul{\gO},0G$\\
   5           &$\bmD^0_1$&2  & 4&$0\ul{\gS},0G$\\
   6           &$\bmC_3$  &3  &12&$1\ul{\gG},1\ul{K},1\ul{L},5G$\\
   $\mathbf{7}$&$\bmC_4$  &4  & 8&$1\ul{\gG},1X,1\ul{S},7G$\\
   $\mathbf{8}$&$\bmD^1_2$&2  & 4&$1\ul{\gG},1\ul{X},1\ul{Y},1\ul{S},3\gD,3\Phi,3\gO,3\gL,1G$\\
   $\mathbf{9}$&$\bmD^0_2$&2  & 4&$1\ul{\gG},1Y,1\ul{S},3\gS,3\Xi,0G$\\
  $\mathbf{10}$&$\bmC_6$&6    &12&$1\ul{\gG},1M,1K,11G$\\
  11           &$\bmD^1_3$&2,3&12&$1\ul{\gG},1K,0\gD,11\Phi,0G$\\
  12           &$\bmD^0_3$&2,3&12&$1\ul{\gG},1\ul{K},1\ul{L},11\gL,11\gS,11\Xi,1G$\\
  $\mathbf{13}$&$\bmD_4$  &4  & 8&$1\ul{\gG},1X,1\ul{S},7\gD,7\gL,7\gS,1G$\\
  $\mathbf{14}$&$\bmD_6$  &6  &12&$1\ul{\gG},1M,1K,11\gD,11\gL,11\gS,1G$\\
  \hline
\end{tabular}\end{table}

{\Briln} zone of the \emph{layer} groups is 2-dimensional (the lattice space is perpendicular to $z$-axis), and  the isogonal groups are  the axial point groups with the crystallographic orders of principle axis: $n=1,2,3,4,6$. As the horizontal mirror plane acts as identity, effective action of these point groups is reduced to
$\bmC_n$, $\bmC_{2n}$, $\bmC_{n}$,  $\bmD_n$, $\bmD_n$, $\bmD_{2n}$ and $\bmD_{n}$, respectively.  For $q>1$ the group $\bmC_q$ in the interior of $\mho$ fixes only $\gG$-point; thus this is the only interior fixed and $\bgg$-point for the layer groups with nontrivial principle axis rotations. When $q=1$, $\bmC_1$ fixes the whole $\mho$; if it is the only effective element of $\bmP$, then $\mho^{\bgg}=\mho_\bmG$, and otherwise the fixed points are within axis (of the group $\bmD_1$) or vertical mirror plane (for $\bmC_{1\textrm{v}}$)). Winding numbers of the axis of $\bmD_n$ give altogether 14 types of irreducible domains~\cite{NLSite}. Here we present data for each of them. For each ID the least common multiple $d$ of the  generators of the associated (double) groups is calculated, and the ID partition into $d$ equivalent parts is found (Fig.~\ref{FIDs2}). Then the  number of the remaining points for each stratum of ID (Table~\ref{TIDs}) is found.  All the integrals appearing in the calculation of  strata determinants of IDs 1, 3, 4 and 5 (when $\gamma$ points are not isolated) are equal to 1. These data suffice to calculate all the strata determinant.

\subsubsection{Line groups}

\squeezetable
\begin{table}
\caption{\label{TLGs} Thirteen families of the line groups. For each family (F) in the column Type H or L is for helical or linear factorization, $S$ or $N$ for (non)symmorphic groups and $+$ and $-$ for the colorless groups with/without $z$-reversing elements (grey group families are $-$), which corresponds to ID ($(-\pi,\pi]$ or $[0,\pi]$.  Then follow columns $n$ (orders of the isogonal group generators)and $d$ (division order  of the generic stratum LCM of the orders of the double group generators)/ Finally,  number $c$ of contributing points for strata $S$ (underlined if $\bgg$-point) are listed for colorless (C) and grey (G) groups.  $x=4n-1$.}
\begin{tabular}{|cc|cc|cc|}
  \hline
  F &Type&$n$&$d$&C: $cS,c\ul{\ggg}$&G: $cS,c\ul{\ggg}$\\
  1 &H\,S\,$+$&$?,n   $&$? $&$?\ul{G}               $&$1\ul{\gG},xG,1\ul{\Pi}$\\
  2 &L\,S\,$-$&$2n    $&$4n$&$1\ul{\gG},xG,1\ul{\Pi}$&$1\ul{\gG},xG,1\ul{\Pi}$\\
  3 &L\,S\,$-$&$n,2   $&$4n$&$1\ul{\gG},xG,1\ul{\Pi}$&$1\ul{\gG},xG,1\ul{\Pi}$\\
  4 &L\,N\,$-$&$2n,n,2$&$4n$&$1\ul{\gG},xG,1\Pi     $&$1\ul{\gG},xG,1\Pi     $\\
  5 &H\,S\,$-$&$?,n,2 $&$? $&$1\ul{\gG},?G,1\ul{\Pi}$&$1\ul{\gG},xG,1\ul{\Pi}$\\
  6 &L\,S\,$+$&$n,2   $&$4n$&$x\ul{G}               $&$1\ul{\gG},xG,1\ul{\Pi}$\\
  7 &L\,N\,$+$&$n,2   $&$4n$&$x\ul{G}               $&$1\ul{\gG},xG,1\Pi     $\\
  8 &L\,N\,$+$&$2n,n,2$&$4n$&$x\ul{G}               $&$1\ul{\gG},xG,1\Pi     $\\
  9 &L\,S\,$-$&$2n,n,2$&$4n$&$1\ul{\gG},xG,1\ul{\Pi}$&$1\ul{\gG},xG,1\ul{\Pi}$\\
  10&L\,N\,$-$&$2n,n,2$&$4n$&$1\ul{\gG},xG,1\Pi     $&$1\ul{\gG},xG,1\Pi     $\\
  11&L\,S\,$-$&$n,2   $&$4n$&$1\ul{\gG},xG,1\ul{\Pi}$&$1\ul{\gG},xG,1\ul{\Pi}$\\
  12&L\,N\,$-$&$n,2   $&$4n$&$1\ul{\gG},xG,1\ul{\Pi}$&$1\ul{\gG},xG,1\Pi     $\\
  13&L\,N\,$-$&$2n,n,2$&$4n$&$1\ul{\gG},xG,1\Pi     $&$1\ul{\gG},xG,1\Pi     $\\
  \hline
\end{tabular}\end{table}

There are 13 infinite families of the line groups\cite{YILG}. We assume periodicity (translational or helical) along $z$-axis. Therefore, principle and helical rotations (around $z$-axis) and vertical mirror planes act in \Briln's zone  as identity, and the isogonal groups $\bmC_n$ and $\bmC_{n\mathrm{v}}$  act as $\bmC_1$ (identity only) giving $\mho_\bmG=\mho$, while other ones as $\bmD_1$ (identity and a $z$-reversal operation), with $\mho_\bmG=[0,\pi/a]$ stratified into interval $G=(0,\pi)$ and points $\gG$ and $\Pi$ (for $k=0,\pi$, respectively).
In the first case (families 1,6,7,8) $\mho^{\bgg}=\mho$, since  the whole BZ is fixed, while \eqref{EgamaRay} is  satisfied automatically. In other families due to the $z$-reversing symmetries, the fixed points are $\gG$ and $\Pi$; \eqref{EgamaRay} is satisfied for $\gG$, and for $\Pi$ in the symmorphic families 2, 3, 5 (helical axis used instead of the translational group, with helical QNs), 9, 11, and nonsymmorphic 10 and 12. In nonsymmorphic families 4 and 13, the only $\bgg$-point is $\gG$. When time reversal is added (grey groups), all the irreducible domain are $[0,\pi]$; $\gG$ is $\ggg$-point in all groups, while $\Pi$ only in symmorphic; in double grey groups all the determinant representations are unit.

\end{document}